\documentclass[12pt]{article}            		

\usepackage[top=1in, bottom=1in, left=1in, right=1in]{geometry}
\usepackage{authblk}
\usepackage{overcite}
\usepackage{graphicx}
\usepackage{subfig}
\usepackage[version=3]{mhchem}
\usepackage{amsmath}
\usepackage[numbers,super,sort&compress,comma]{natbib}

\usepackage{xr}
\title{An Electrostatic Spectral Neighbor Analysis Potential (eSNAP) for Lithium Nitride}
\author[]{Zhi Deng}
\author[]{Chi Chen}
\author[]{Xiang-Guo Li}
\author[]{Shyue Ping Ong\thanks{Email: ongsp@eng.ucsd.edu}}
\affil[]{\textit{Department of NanoEngineering, University of California San Diego, 9500 Gilman Dr, Mail Code 0448, La Jolla, CA 92093-0448, United States}}
\date{}

\begin{document}
\maketitle

\begin{abstract}
Machine-learned interatomic potentials based on local environment descriptors represent a transformative leap over traditional potentials based on rigid functional forms in terms of prediction accuracy. However, a challenge in their application to ionic systems is the treatment of long-ranged electrostatics. Here, we present a highly accurate electrostatic Spectral Neighbor Analysis Potential (eSNAP) for ionic $\alpha$-\ce{Li3N}, a prototypical lithium superionic conductor of interest as a solid electrolyte or coating for rechargeable lithium-ion batteries. We show that the optimized eSNAP model substantially outperforms traditional Coulomb-Buckingham potential in the prediction of energies and forces, as well as various properties, such as lattice constants, elastic constants and phonon dispersion curves. We also demonstrate the application of eSNAP in long-time, large-scale Li diffusion studies in \ce{Li3N}, providing atomistic insights into measures of concerted ionic motion (e.g., the Haven ratio) and grain boundary diffusion. This work aims at providing an approach to developing quantum-accurate force fields for multi-component ionic systems under the SNAP formalism, enabling large scale atomistic simulations for such systems. 
\end{abstract}

\section*{Introduction}

A potential energy surface (PES) that yields potential energy of a system of atoms with given atomic coordinates is the fundamental enabler for atomistic simulation methods. In principle, \textit{ab initio} or first principles methods that solve the Schr\"odinger equation, typically some approximation within the Kohn-Sham density functional theory (DFT) framework,\cite{Kohn1964,Sham1965} can be applied to directly calculate the PES. While such methods are highly accurate and transferable across diverse chemistries and bonding types, their high computational cost limit their application in molecular dynamics (MD) simulations to relatively small and simple systems containing up to a few hundreds of atoms and sub-nanosecond time scales. Empirical interatomic potentials, on the other hand, are a much cheaper alternative. The functional form of these potentials are drastically simplified with only a few fitting parameters to satisfy physical considerations.\cite{Buckingham1938,Daw1984} However, the accuracy of the empirical potentials is necessarily limited by the approximations made in selecting the functional form, which are generally not transferable to another system with different bonding types. 

In recent years, an alternative approach has gained popularity in constructing interatomic potentials with improved transferability.\cite{Behler2007,Artrith2011,Bartok2010a,Bartok2013b,Thompson2015a,Shapeev2016a} In this approach, the atomic coordinates are featurized using local environment descriptors that are invariant to translations, rotations and permutations of homo-nuclear atoms, and are differentiable and unique.\cite{Behler2011a,Bartok2013b} A machine learning model is then trained to map the structural features to data (energies, forces, etc.) from first principles calculations. Such potentials have been demonstrated to achieve accuracy close to first principles methods at much lower computational costs.\cite{Behler2007,Bartok2010a,Thompson2015a,Shapeev2016a}

The coefficients of the bispectrum of local atomic density were first applied in the Gaussian approximation potential by \citet{Bartok2010a} Thompson et al. later showed that a linear model of bispectrum coefficients from the lowest order - the so-called Spectral Neighbor Analysis Potential (SNAP) - can accurately reproduce DFT energies and forces as well as a variety of calculated properties (e.g., elastic constants and migration barrier for screw dislocations) in bcc Ta and W.\cite{Thompson2015a,Wood2018} More recently, the current authors have extended the SNAP formalism to bcc Mo, fcc Ni and Cu, and the binary fcc Ni-bcc Mo alloy systems and showed that it outperforms traditional embedded atom method (EAM) and modified EAM potentials across a wide range of properties. \cite{Chen2017b,Li2018c} Thus far, SNAP models have mainly been developed for metallic systems.

\begin{figure}[htp]
\centering
\includegraphics[width=0.8\textwidth]{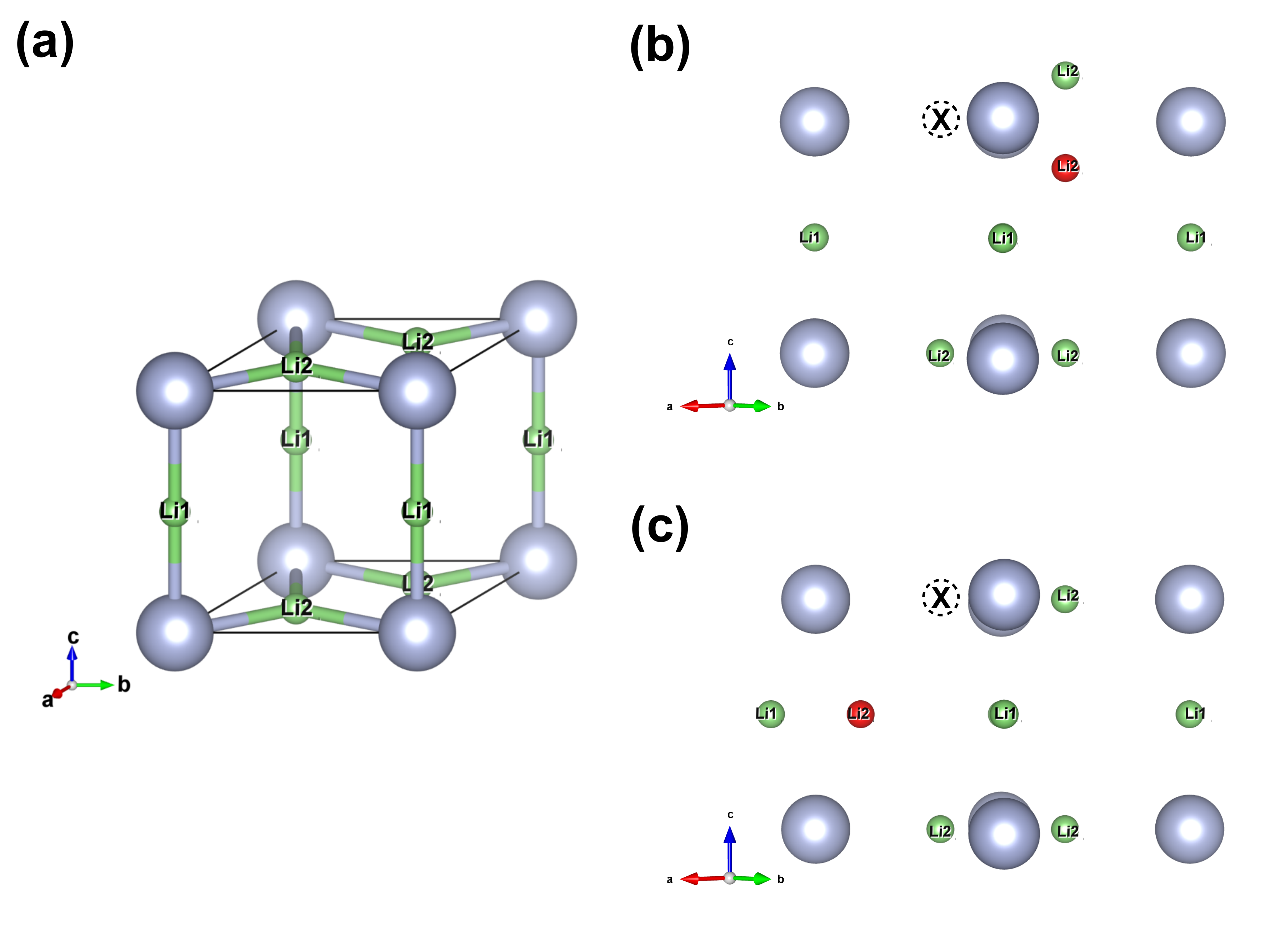}
\caption{(a) Unit cell of $\alpha$-\ce{Li3N} (space group: $P6/mmm$). Green: Li; grey: N. (b) Intra-planar and (c) inter-planar Frenkel defect configuration. The vacancy (white dashed circle X) forms at Li1 site for both cases, while the interstitial (red) forms at Li2 and Li1 site for intraplanar and interplannar configurations, respectively. }
\label{fig:crystal}
\end{figure}

For ionic systems, a common strategy in constructing interatomic potentials is to incorporate long-ranged electrostatic interactions (e.g., through the use of the Ewald summation) on top of energy model. This has been done for both traditional empirical models\cite{Lewis1985,Lee2016} as well as modern local atomic environment descriptor-based potentials (e.g., GAP for the mixed ionic-covalent \ce{GaN}\cite{Bartok2010a} and neural network potential for \ce{ZnO}\cite{Artrith2011}). In this work, we develop a highly-accurate electrostatic SNAP (eSNAP) model for ionic $\alpha$-\ce{Li3N} (see Figure \ref{fig:crystal}(a)). $\alpha$-\ce{Li3N} is one of the earliest lithium superionic conductors ever reported,\cite{Boukamp1978} and remains a promising solid electrolyte/anode coating candidate today due to its stability against Li metal.\cite{Zhu2017c,Li2018b} A highly accurate potential model for $\alpha$-\ce{Li3N} would enable large-scale, long-time-scale  diffusion studies of this highly important prototypical lithium conductor, as well as serve as a platform in which to develop similar potentials for more complex systems.

\section{Results}

\subsection*{Optimized model parameters}

In our proposed electrostatic SNAP (eSNAP) model, we write the total potential energy $E_p$ as the sum of the electrostatic contributions and the local (SNAP) energy due to the variations in atomic local environments (SNAP), as follows:
\begin{eqnarray}
E_p & = & \gamma E_{el} + E_{SNAP}\label{esnapeqn}\\
\mathbf{F}_j & = & -\bigtriangledown_jE_p = -\gamma\bigtriangledown_jE_{el}-F_{j, SNAP}
\end{eqnarray}
where $E_{el}$ and $E_{SNAP}$ are the electrostatic energy computed using the Ewald summation approach\cite{Ewald1921} and the energy from SNAP, respectively, and $\gamma$ is an effective screening prefactor for electrostatic interactions.  An iterative procedure was developed to fit all model parameters using total energies and forces from DFT calculations until the training and test errors are converged (see Methods for details).

For \ce{Li3N}, we calculated the electrostatic energy by assigning formal charges 1 and -3 to Li and N, respectively. For highly ionic $\alpha$-\ce{Li3N}, we find that assigning formal charges, with screening accounted for via a fitted parameter ($\gamma$ in Equation \ref{esnapeqn}), results in a simpler, more stable potential model than variable charge models such as the charge equilibration (QEq)\cite{Rappe1991} method. The narrow charge distribution of Li atoms from Bader analysis (See Figure \ref{fig:bader}) also supports the usage of fixed charge.  The final hyperparameters and coefficients for the optimized eSNAP model are given in Table \ref{tab:param_coeff}. The optimized effective screening parameter $\gamma$ is 0.057.

\begin{table}[htp]
\centering
\caption{Final hyperparameters and coefficients of SNAP.}
\label{tab:param_coeff}
{\footnotesize
\begin{tabular}{rrrrrr}
\hline
 & & & & Li ($w=0.1, R=2.0$) & N ($w=-0.1, R=2.8$)\\
$k$ & $2j_1$ & $2j_2$ & $2j$ & $\beta_{{\rm Li}, k}$ & $\beta_{{\rm N}, k}$ \\
\hline
0 & & & & -41.973239307589510000 & 5.070489350565292000 \\
1 & 0 & 0 & 0 & -0.006975291543708552 & 1.456429449741217000 \\
2 & 1 & 0 & 1 & 1.943443830363772200 & -0.741751895178898800 \\
3 & 1 & 1 & 2 & 1.943960665209158600 & 1.391487252718133000 \\
4 & 2 & 0 & 2 & 1.896372127297504700 & -0.057993919814256490 \\
5 & 2 & 1 & 3 & 0.818415025331073900 & 4.490489555500009000 \\
6 & 2 & 2 & 2 & -0.115525745627534120 & 1.038809274414984000 \\
7 & 2 & 2 & 4 & 0.024662060558673055 & 2.017342084202995000 \\
8 & 3 & 0 & 3 & 0.645913664832655000 & 0.297752638364790700 \\
9 & 3 & 1 & 4 & 0.029473213939390678 & 2.234672087302177000 \\
10 & 3 & 2 & 3 & -0.835569404051590300 & 0.864212802963693300 \\
11 & 3 & 2 & 5 & 0.260421577039490040 & 2.027941600818144300 \\
12 & 3 & 3 & 4 & -0.531877413286821100 & 0.534495092798581400 \\
13 & 3 & 3 & 6 & 0.297955449169754300 & 0.299436757239075040 \\
14 & 4 & 0 & 4 & 0.100238669102941730 & -0.149453371719054230 \\
15 & 4 & 1 & 5 & -0.569297582876690400 & 0.541280152554589700 \\
16 & 4 & 2 & 4 & -0.683610621813019800 & 0.181446484735933700 \\
17 & 4 & 2 & 6 & 0.076839221845524830 & 0.324534968809946260 \\
18 & 4 & 3 & 5 & 0.157477937603063170 & -0.112365239839474810 \\
19 & 4 & 4 & 4 & 0.245396146771298870 & 0.143146336052880570 \\
20 & 4 & 4 & 6 & 0.218995247596486900 & 0.008000929122986318 \\
21 & 5 & 0 & 5 & -0.203265539796318980 & -0.191267051611771400 \\
22 & 5 & 1 & 6 & -0.232726613086012750 & -0.167152854094159280 \\
23 & 5 & 2 & 5 & -0.352838031703020600 & -0.369717312625315600 \\
24 & 5 & 3 & 6 & 0.112930317776087000 & -0.123035057771956320 \\
25 & 5 & 4 & 5 & 0.537951697554698000 & 0.559745649003179100 \\
26 & 5 & 5 & 6 & 0.082284784064962830 & 0.290496672932174600 \\
27 & 6 & 0 & 6 & -0.178966722576012600 & -0.029092905917972420 \\
28 & 6 & 2 & 6 & -0.283676416415676500 & -0.180134374249985380 \\
29 & 6 & 4 & 6 & 0.089012939842931950 & 0.252331183046023500 \\
30 & 6 & 6 & 6 & 0.044042461635336136 & 0.008318824055866198 \\
\hline
\end{tabular}}
\end{table}

\subsection*{Energy and Force Prediction}

\begin{figure}[htp]
\centering
\includegraphics[width=\textwidth]{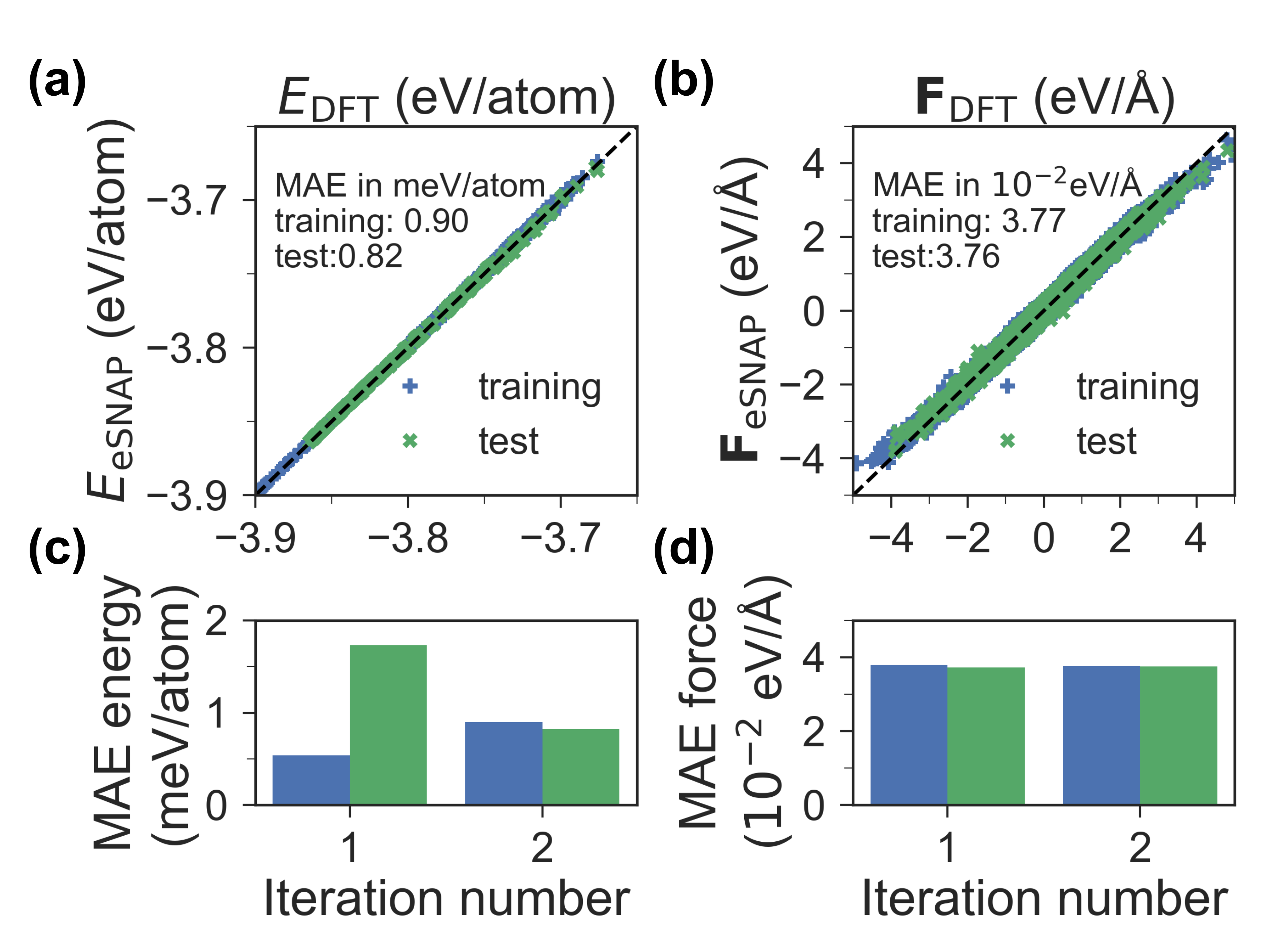}
\caption{Energy and force prediction errors for eSNAP. Comparisons between DFT and eSNAP predictions for (a) energies and (b) forces on both training and test dataset in the final iteration. Convergence of the test and training MAEs for (c) energies and (d) forces with iteration number.}
\label{fig:energy+force}
\end{figure}

Figures \ref{fig:energy+force}a-b shows the comparison between DFT calculated and eSNAP predicted energies and forces on both training and test dataset in the final iteration. Both energy and force predictions agree well with those from DFT calculations, indicating the eSNAP model has successfully captured the fundamental relationship between atomic environment and potential energy/atomic forces. The MAEs on energies and forces reached convergence after only two iterations, as shown in Figures \ref{fig:energy+force}c-d. In comparison, the MAE between DFT and the Coulomb-Buckingham potential by \citet{Walker1981} on the initial training configuration pool are substantially higher for both energies (22 meV/atom) and forces (0.48 eV/\AA). 

\subsection*{Structural Properties}

\begin{table}[htp]
\centering
\caption{Calculated lattice constants, elastic constants, vacancy migration barriers and Frenkel defect formation energies with DFT, eSNAP and Coulomb-Buckingham potential.\cite{Walker1981} Lattice constants\cite{Rabenau1976} and elastic constants\cite{Kress1980} from experimental measurements are also listed for comparison.}
\label{tab:structural}
\begin{tabular}{ccrrrr}
\hline
 & & DFT & eSNAP & Coul-Buck & Exp. \\
\hline
Lattice constant & $a$ & 3.641 & 3.641 & 3.528 & 3.648 \\
(\AA) & $c$ & 3.874 & 3.872 & 3.628 & 3.875 \\
\hline
Elastic constant & $c_{11}$ & 123 & 116 & 165 & 114 \\
(GPa) & $c_{33}$ & 137 & 144 & 193 & 118 \\
 & $c_{44}$ & 17 & 17 & 19 & 17 \\
 & $c_{66}$ & 48 & 39 & 53 & 38 \\
\hline
Defect formation energy & Intra-planar & 0.60 & 0.64 & 0.44 & \\
(eV) & Inter-planar & 0.51 & 0.63 & 0.46 & \\
\hline
Defect migration barrier & Intra-planar (vacancy) & 0.04 & 0.04 & N/A & \\
(eV) & Inter-planar (interstitial) & 0.22 & 0.09 & N/A & \\
\hline
\end{tabular}
\end{table}

Table \ref{tab:structural} compares the computed physical properties of $\alpha$-\ce{Li3N} with different potential energy surfaces. The lattice constants calculated from eSNAP agree with those from DFT and experiments.\cite{Rabenau1976} The calculated elastic constants from eSNAP also match reasonably well with DFT calculated and experimental values.\cite{Kress1980}  This excellent agreement on structural properties can be expected from the fact that the energies of unit cells with various distortions have been fed to the model with a large sample weight. In comparison, the lattice constants and elastic constants from the Coulomb-Buckingham potential match poorly with both DFT and experimental values, despite the fact that these physical properties were used to determine the potential parameters.\cite{Walker1981} 

We have also calculated the formation energy of Li Frenkel defects and the migration barrier of these defects. We considered two Frenkel configurations where a vacancy is introduced on a Li2 site and the interstitial Li is located at either the Li2 site (intra-planar, Figure \ref{fig:crystal}b) or Li1 site (inter-planar, Figure \ref{fig:crystal}c), and all defect configurations are fully relaxed within each potential. The eSNAP model yields reasonably close formation energy of intra-planar defect to the DFT value, but slightly overestimates the value of inter-planar defect by 0.12 eV. On the other hand, the Coulomb-Buckingham potential underestimates the defect formation energies, likely due to the use of unsatisfactory lattice constants in building the defect configurations. Using the nudged elastic band (NEB) method,\cite{Henkelman2000} we calculated the migration barriers of two types of hops, namely intra-planar (Li2 to Li2) vacancy migration and inter-planar (Li1 to Li2) Li interstitial migration. As shown in Table \ref{tab:structural}, the eSNAP barrier for intra-planar vacancy migration is in good agreement with the DFT barrier, while the eSNAP barrier for inter-planar interstitial migration underestimates the DFT barrier by 0.13 eV. We note that the eSNAP defect formation energies and migration barriers for the dominant intra-planar diffusion direction are reasonably close to the DFT values, while the over-estimation of the inter-planar defect formation energy by eSNAP is compensated by the under-estimation of the vacancy migration barrier. Similar errors and error compensations have been reported in prior non-electrostatic SNAP models on metals such as Mo and Ni.\cite{Chen2017b,Li2018c} On the other hand, we are unable to converge the NEB barriers using the Coulomb-Buckingham potential due to its inability to model the transition states. 

\begin{figure}[htp]
\centering
\includegraphics[width=0.5\textwidth]{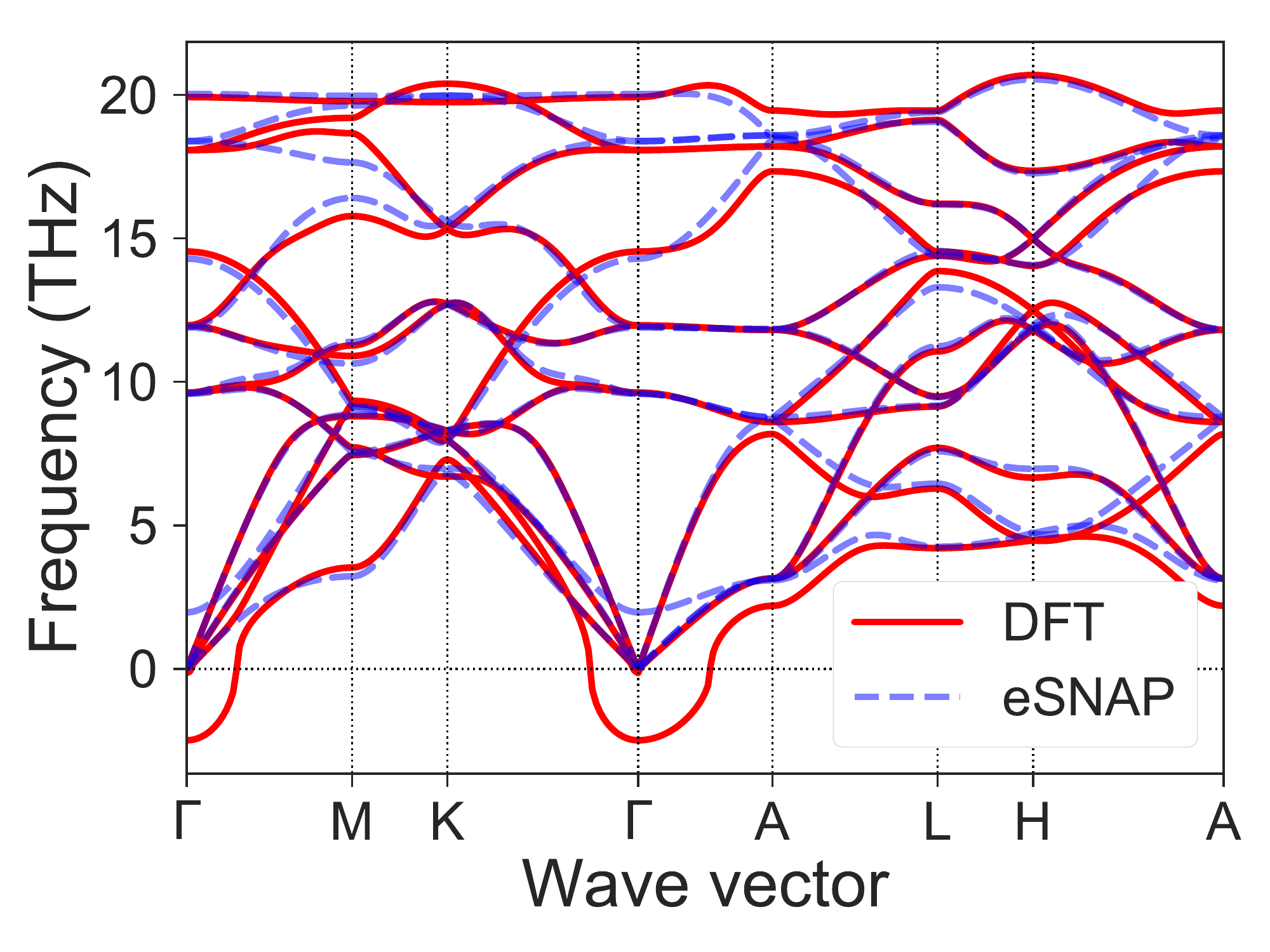}
\caption{Phonon dispersion curves of $\alpha$-\ce{Li3N} calculated from DFT and eSNAP.}
\label{fig:phonon}
\end{figure}

Finally, Figure \ref{fig:phonon} compares the calculated phonon dispersion curves of $\alpha$-\ce{Li3N} from eSNAP with those from DFT calculations. The phonon dispersion curves were calculated using the finite displacement approach on a $3\times3\times3$ supercell as implemented in the phonopy package.\cite{Togo2015} We find that the phonon dispersion curves calculated from eSNAP are in good agreement with that from DFT. The only discrepancy is the imaginary phonon mode at $\Gamma$ point observed in DFT phonon dispersion. According to \citet{Wu2011}, this lattice instability is associated with the vibration of Li2 sites along the $c$ axis, resulting in a more stable phase that is only 0.3 meV/atom lower in energy after displacing Li2 site by $\sim$ 0.1 \AA. This energy difference is well within the energy prediction error of the eSNAP model. We also note that the experimentally measured phonon dispersion curves at room temperature do not exhibit this lattice instability.\cite{Kress1980} In contrast, the phonon dispersion curve calculated from the Coulomb-Buckingham potential show severely overestimated frequencies (Figure \ref{fig:phonon_buck}) due to its unsatisfactory force prediction. 

\subsection*{Bulk diffusion}

MD simulations were performed using the optimized eSNAP to investigate Li diffusion in bulk $\alpha$-\ce{Li3N}. Built from the unit cell with equilibrium volume, the simulation box is a $5\times5\times5$ supercell of bulk $\alpha$-\ce{Li3N} containing 500 atoms. MD simulations were carried out at elevated temperatures from 600 to 1200 K in an NVT ensemble for 1 ns long. 

We first validated the eSNAP potential by comparing the mean square displacement (MSD) and diffusivities obtained from eSNAP MD simulations with those obtained from AIMD simulations at high temperatures (1000 and 1200 K). Runs at lower temperatures were not chosen due to the poor convergence of diffusivity at limited simulation length (40 ps). It should be noted that even though 1200 K is above the melting point of \ce{Li3N}, the lattice did not melt in either AIMD or eSNAP MD during the short period of simulations. As shown in Figure \ref{fig:msd_bulk}, the generally high Li mobility and anisotropic diffusion in $\alpha$-\ce{Li3N} are successfully reproduced with eSNAP MD simulations. The tracer diffusivities (given by the slope of the MSD with respect to time) from eSNAP MD ($1.48\times10^{-4}$ cm$^2$/s at 1000 K, $2.35\times10^{-4}$ cm$^2$/s at 1200 K) are in generally good agreement with those from AIMD ($1.28\times10^{-4}$ cm$^2$/s at 1000 K, $2.16\times10^{-4}$ cm$^2$/s at 1200 K), showing a slight overestimation of about 15\% and 8\% at 1000 K and 1200 K, respectively. 

\begin{figure}[htp]
\centering
\includegraphics[width=0.5\textwidth]{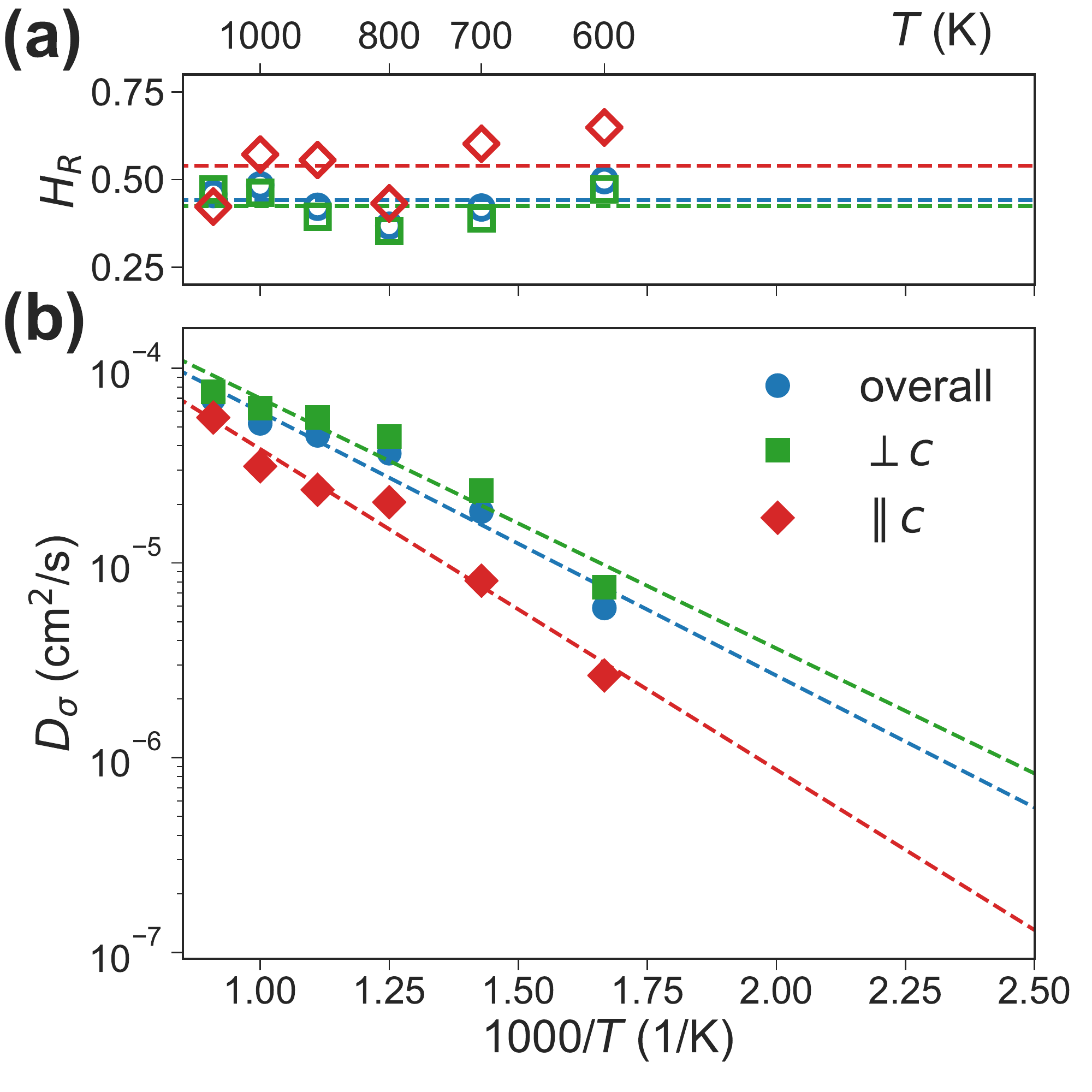}
\caption{(a) Haven ratio and (b) Arrhenius plot for Li charge diffusivity in bulk $\alpha$-\ce{Li3N} obtained from eSNAP MD simulations.}
\label{fig:bulk}
\end{figure}

Beyond tracer diffusivities, the orders of magnitude lower computational cost of the eSNAP relative to DFT affords us the capability to compute the charge diffusivity $D_\sigma$. For each temperature, 100 independent simulations were performed starting from different initial velocities. Diffusivities were obtained by averaging square displacements over all simulations at a particular temperature.  Figure \ref{fig:bulk} plots the predicted Haven ratio and Arrhenius plot for \ce{Li3N} from eSNAP MD simulations. The activation energies, extrapolated room temperature conductivities and average Haven ratio across all temperatures are tabulated in Table \ref{tab:ea_con}. The anisotropic diffusion in $\alpha$-\ce{Li3N} observed experimentally\cite{Alpen1977,Messer1981} is reproduced in many aspects, including the magnitude of diffusivity, activation energy and Haven ratio. The higher diffusivities and lower activation energy in the direction perpendicular to $c$ axis is consistent with the lower Haven ratio found. The activation energy perpendicular to $c$ axis close to the one in single crystal measurement, though the value parallel to $c$ axis is much lower compared with experiments.\cite{Alpen1977} The lower activation energies lead to much higher extrapolated room temperature ionic conductivity for both directions. The Haven ratios obtained from eSNAP MD are reasonably close to the NMR measured values. In comparison, we also performed a similar series of MD simulations with the Coulomb-Buckingham potential \cite{Walker1981}, and the results significantly underestimate the fast ionic conduction in $\alpha$-\ce{Li3N}, and significantly overestimates the Haven ratio. In particularly, the Haven ratio for the direction parallel to the c-axis is computed to be $>$ 1 using the Coulomb-Buckingham potential.

\begin{table}[htp]
\centering
\caption{Comparison of activation energies ($E_a$), room temperature ionic conductivities ($\sigma_{\rm RT}$) and Haven ratio ($H_R$) obtained from eSNAP and Coulomb-Buckingham MD\cite{Walker1981} with single crystal dc conductivity\cite{Alpen1977} and NMR\cite{Messer1981} measurements.}
\label{tab:ea_con}
\begin{tabular}{llrrr}
	\hline
	 & & eSNAP & Coul-Buck\cite{Walker1981} & Exp. \\
	\hline
	$E_a$ & $\perp c$ & 0.255 & 0.422 & 0.290\cite{Alpen1977}  \\
	(eV) & $\parallel c$ & 0.327 & 0.597 & 0.490\cite{Alpen1977}  \\
	 & overall & 0.269 & 0.432 & \\
	\hline
	$\sigma_{\rm RT}$ & $\perp c$ & 29.6 & 0.228 & 1.20\cite{Alpen1977}  \\
	(mS/cm) & $\parallel c$ & 2.32 & 0.0004 & 0.01\cite{Alpen1977}  \\
	 & overall & 17.3 & 0.129 & \\
	\hline
	Avg. $H_R$ & $\perp c$ & 0.42 & 0.47 & 0.3\cite{Messer1981} \\
	 & $\parallel c$ & 0.54 & 1.2 & 0.5\cite{Messer1981} \\
	 & overall & 0.44 & 0.51 & \\
	\hline
\end{tabular}
\end{table}

\subsection*{Grain boundary diffusion}

\begin{figure}[htp]
\centering
\includegraphics[width=0.7\textwidth]{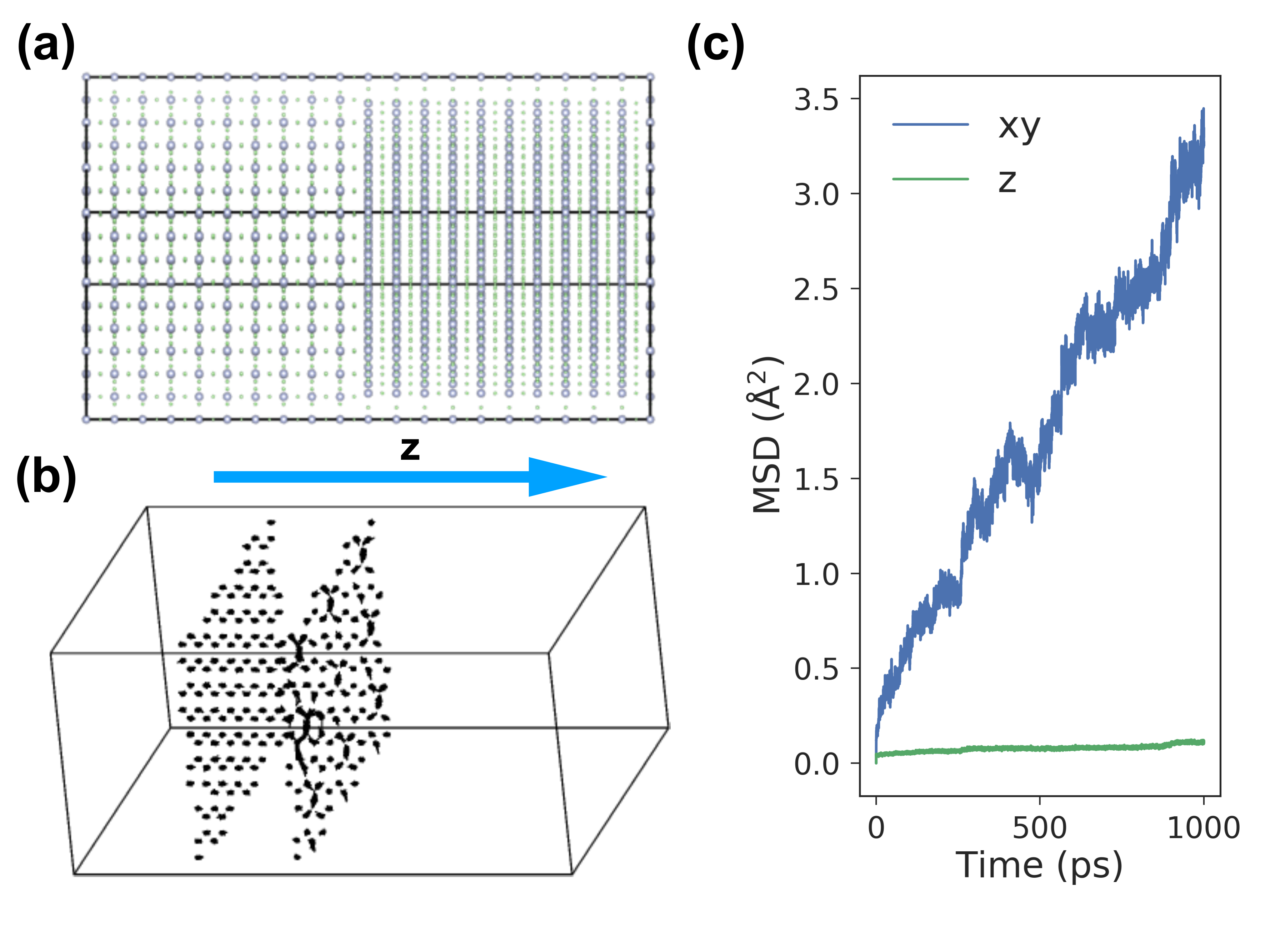}
\caption{(a) Constructed simulation boxes with twist $\Sigma7$ [0001] grain boundaries. (b) Trajectories for selected Li ions in the box with twist grain boundaries in 0.5 ns. Li ions on the left lie in the bulk region, and the ones on the right are close to one of the grain boundaries. (c) MSD (by components) vs. time plot for Li ions located at grain boundaries in the simulation box with twist grain boundaries. The z direction is perpendicular to the grain boundaries.}
\label{fig:gb}
\end{figure}

To investigate GB diffusion, we first computed the grain boundary (GB) energies of two low $\Sigma$ twist grain boundary (GB) configurations - $\Sigma4$ [1000] and $\Sigma7$ [0001]. Both configurations are fully relaxed using DFT and eSNAP. The eSNAP-calculated GB energies for twist $\Sigma4$ [1000] and $\Sigma7$ [0001] GBs are 1.41 and 0.85 J m$^-2$, respectively, in good agreement with the DFT values of 1.64 J m$^-2$ and 0.86 J m$^-2$, respectively. 

The lower-energy twist $\Sigma7$ [0001] GB is then used in large scale diffusion studies, as shown in Figure \ref{fig:gb}. The simulation box (Figure \ref{fig:gb}a) contains 5,040 atoms in total. Due to the periodic boundary conditions, two grain boundaries separated by $10\times$ lattice vector $c$ present in the box. NVT MD simulations were carried out at 300 K, with thermalization carried out for 30 ps  followed by production simulations of 1 ns. We find that the mean square displacement (MSD) of Li atoms within the GB plane are much higher than in the bulk region, as shown in Figure \ref{fig:gb}b and \ref{fig:gb}c, and there are few migration events occurring between the GB layer and the bulk layers. From the MSD, we estimate the 2D Li self-diffusivity within the twist grain boundary to be $7.09\times10^{-8}$ cm$^2$/s, about 3 times of extrapolated total value ($2.24\times10^{-8}$ cm$^2$/s in 3D) in the bulk at 300 K. These results indicate that grain boundaries may provide a rapid pathway for Li diffusion in $\alpha$-\ce{Li3N}.

\section*{Discussion}

In this work, we demonstrate that modern potentials based on local environment descriptors such as the Spectral Neighbor Analysis Potential (SNAP) can be adapted for ionic systems by incorporating long-range electrostatics.  

The introduction of $\gamma$ as a hyperparameter offers more flexibility to the potential model in order to achieve higher predictive power. Physically, $\gamma$ can be interpreted as the inverse of dielectric constant. Indeed, the optimized value of $\gamma$ is 0.057, which implies an effective dielectric constant of 17.5, reasonably close to the experimental dielectric of $\alpha$-\ce{Li3N} of 14.\cite{Wahl1978} We note that while the experimentally measured dielectric constant could have been provided as an input to model development, the goal of this effort is to develop a general approach to training eSNAP models for materials, some of which may not have measured dielectric constants. We have also attempted to fit a regular SNAP model for \ce{Li3N} without the use of electrostatic interactions, but using a larger cutoff radius of 8 \AA to allow the model to learn screened electrostatic interactions. The resulting SNAP model has significant higher MAEs in energies and forces of 2.3 meV/atom and 0.15 eV \AA$^{-1}$, respectively.

Unlike earlier works where the sample weights are treated as hyperparameters optimized towards structural properties (lattice constants, elastic constants, etc.),\cite{Chen2017b,Li2018c} we used fixed sample weights in linear regression as the different scales between energies and forces are unified by using standardized Z-scores as targets. Sample weight assignment then effectively becomes an exercise in assigning importance of matching various computed properties from DFT. Note that reproducing energetic calculations where atoms are relaxed remains a challenge, as eSNAP could not distinguish the difference of defect formation energies in different Frenkel defect configurations. 

It should be noted that the focus of the current eSNAP model is on reproducing the energies and forces on solid-phase $\alpha$-\ce{Li3N} for the purposes of scaling MD simulations beyond the limited simulation cells and time scales in AIMD for diffusion studies. As such, the training structures were selected previously for this purpose and no attempt was made to include a broad diversity of training structures from different polymorphs of \ce{Li3N}, liquid configurations, etc. in the training pool. 

Our choice of the SNAP approach is motivated by its simple linear form and its efficient implementation in the widely-available open-source LAMMPS Molecular Dynamics Simulator.\cite{Plimpton1995} Though the mean absolute errors (MAEs) of linear SNAP may not be as low as those achieved using other regression models and descriptors,\cite{Bartok2013b,Szlachta2014} its efficiency and low training data requirements are the decisive factors for our choice. In terms of scaling performance, we tested the running time of a 1000-step MD simulation with various system sizes (500-500000 atoms). Despite the $O(N\log N)$ time complexity of Ewald summation, eSNAP generally shows linear scaling performance (Figure \ref{fig:scale}), presumably governed by the time-consuming bispectrum coefficient calculations. 

Finally, we applied the eSNAP model to conduct long time-scale ($\sim$ 1 ns) simulations of complex models (500-5000 atoms) of $\alpha$-\ce{Li3N}. We report the Haven ratio of $\alpha$-\ce{Li3N} by directly calculating charge diffusivity and show that grain boundaries may provide faster diffusion pathways (relative to bulk). The calculation of charge diffusivity, which is difficult to converge in \textit{ab initio}-based approaches, enables us to compute much more reliable estimates of the anisotropic diffusivities of $\alpha$-\ce{Li3N}. Interestingly, though we find that conductivity in the $c$-crystallographic direction is in general slower than the $ab$ plane, the value is only one order of magnitude lower, contrary to single crystal measurements.\cite{Alpen1977}. \citet{Li2018b} have recently grown pinhole-free \ce{Li3N} nanofilms as a protective layer on Li metal anodes by flowing nitrogen gas. A critical design requirement is that the conductivity of Li in the [001] direction is sufficiently high. \citet{Li2018b} measured conductivities of up to 0.5 mS/cm, which is in good agreement with our predictions and in disagreement with prior experiments and simulations with the Coulomb-Buckingham potential. It should be emphasized that the conductivity of $\sim$0.01 mS/cm in the $c$ direction reported in previous work\cite{Alpen1977} would lead to a highly resistive, low-performing coating. We hope that further careful experiments in the near future may shed further light on these discrepancies in anisotropic diffusivities between different experiments and computational simulations on this highly important lithium conductor. 

\section*{Methods}

\subsection*{Electrostatic SNAP (eSNAP) Model}

The atomic environment around atom $i$ at coordinates $\mathbf{r}$ can be described by its atomic neighbor density $\rho_i(\mathbf{r})$ with the following equation\cite{Bartok2010a,Thompson2015a}:
\begin{equation}
\rho_i(\mathbf{r}) = \delta(\mathbf{r}) + \sum_{r_{ii'} < R_{ii'}}f_c(r_{ii'})w_{i'}\delta(\mathbf{r} - \mathbf{r_{ii'}}),
\end{equation}
where $\mathbf{r_{ii'}}$ is the vector joining the coordinates of central atom $i$ and its neighbor atom $i'$, the cutoff function $f_c$ ensures that the neighbor atomic density decays smoothly to zero at cutoff radius $R_{ii'}$, and the dimensionless neighbor weights $w_{i'}$ distinguish atoms of different types. This density function can be expanded as a generalized Fourier series in the 4D hyper-spherical harmonics $U^j_{m,m'}(\theta, \phi, \theta_0)$ as follows:
\begin{equation}
\rho_i(\mathbf{r}) = \sum^{\infty}_{j=0, \frac{1}{2}, ...}\sum^j_{m=-j}\sum^j_{m'=-j}u^j_{m,m'}U^j_{m,m'}(\theta, \phi, \theta_0),
\end{equation}
where the coefficients $u^j_{m,m'}$ are given by the inner product $\langle U^j_{m,m'}|\rho\rangle$. The bispectrum coefficients are then given as:
\begin{equation}
B_{j_1,j_2,j} = \sum^{j_1}_{m_1,m'_1=-j_1}\sum^{j_2}_{m_2,m'_2=-j_2}\sum^{j}_{m,m'=-j}\left(u^j_{m,m'}\right)^*H\substack{jmm'\\j_1m_1m'_1\\j_2m_2m'_2}u^{j_1}_{m_1,m'_1}u^{j_2}_{m_2,m'_2},
\end{equation}
where the constants $H\substack{jmm'\\j_1m_1m'_1\\j_2m_2m'_2}$ are coupling coefficients.

In the original formulation of the non-ionic SNAP model,\cite{Thompson2015a} the energy and forces are expressed as a linear function of the bispectrum coefficients, as follows: 

\begin{eqnarray}
E_{SNAP} =& \sum_{\alpha}\left(\beta_{\alpha,0}N_\alpha + \sum_{k=\{j_1,j_2,j\}}\beta_{\alpha,k}\sum^{N_\alpha}_{i=1}B_{k,i}\right)\\
\mathbf{F}_{j,SNAP} =& - \sum_\alpha\sum_{k=\{j_1,j_2,j\}}\beta_{\alpha,k}\sum^{N_\alpha}_{i=1}\frac{\partial B_{k,i}}{\partial \mathbf{r}_j}.
\end{eqnarray}
where $\alpha$ is the chemical identity of atoms, $N_\alpha$ is the total number of $\alpha$ atoms in the system, and $\beta_{\alpha,k}$ are the coefficients in the linear SNAP model for type $\alpha$ atoms. 

\begin{figure}[htp]
\centering
\includegraphics[width=0.5\textwidth]{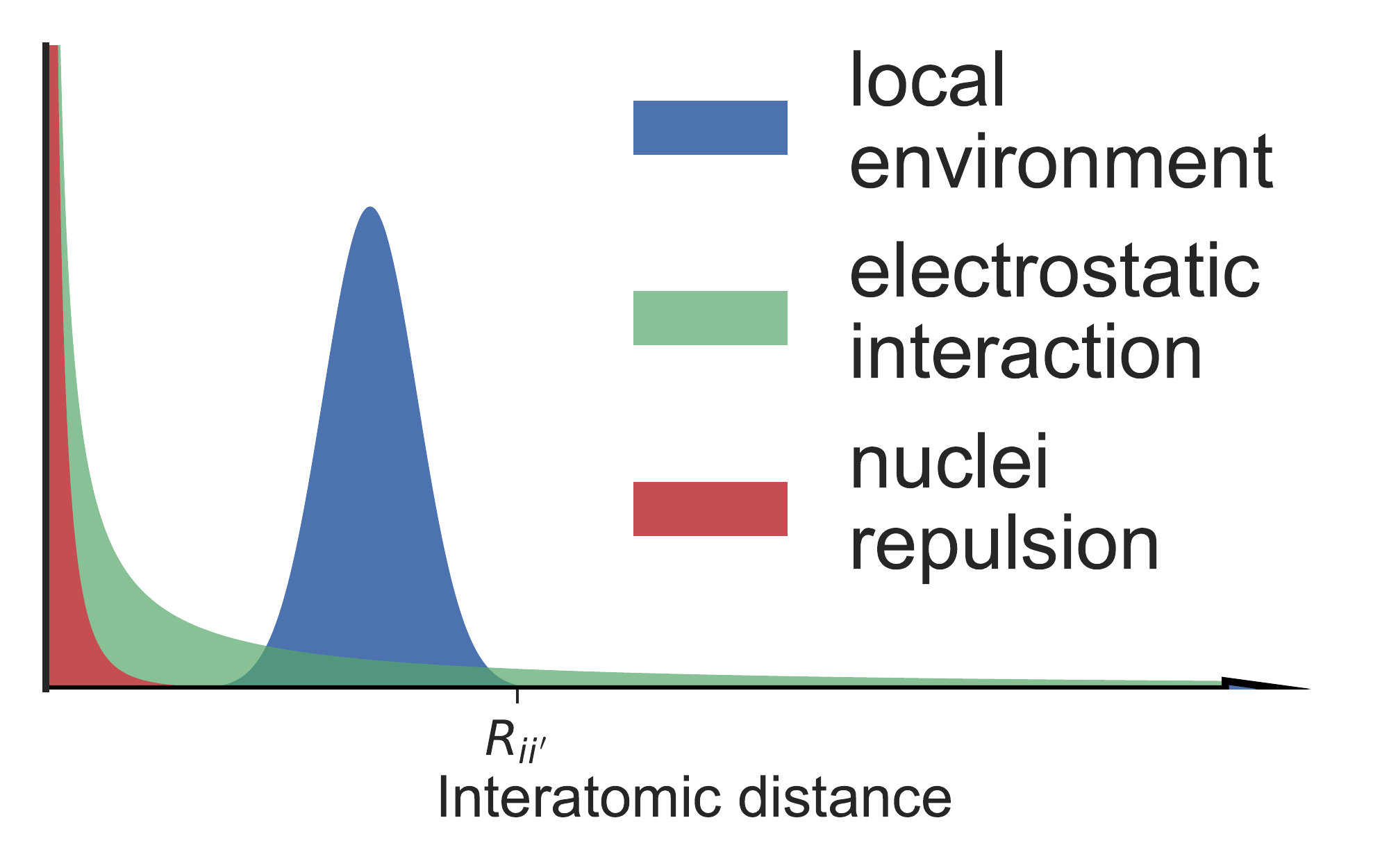}
\caption{Schematic of energy contributions vs. interatomic distances in ionic systems. $R_{ii'}$ denotes the cutoff radius in considering contributions from local environment.}
\label{fig:schema}
\end{figure}

For ionic systems, electrostatic interactions spanning in the entire range of inter-atomic distances are indispensable in the construction of energy model due to the long-range tail beyond the cutoff distance for local environment description (see Figure \ref{fig:schema}). In our proposed electrostatic SNAP (eSNAP) model, we write the total potential energy as the sum of the electrostatic contributions and the local (SNAP) energy due to the variations in atomic local environments (SNAP), as follows:
\begin{eqnarray}
E_p & = & \gamma E_{el} + E_{SNAP}\\
\mathbf{F}_j & = & -\bigtriangledown_jE_p = -\gamma\bigtriangledown_jE_{el}-F_{j, SNAP}
\end{eqnarray}
where $E_{el}$ is the electrostatic energy computed using the Ewald summation approach\cite{Ewald1921} and $\gamma$ is an effective screening prefactor for electrostatic interactions. The coefficients ($\gamma$ and $\beta$) can be solved by fitting the linear model to total energies and forces from DFT calculations. 

In addition, nuclei repulsion emerge at extremely short inter-atomic distances. In this work, the Ziegler-Biersack-Littmark (ZBL) potential is used to account for short-ranged nuclei repulsion.\cite{Ziegler1985} To ensure that the fitting process captures the relevant relationship between the bispectrum coefficients and the DFT energies and forces, the cutoff distances of ZBL were chosen to be short enough ($R_i=1.0$ \AA, $R_o=1.5$ \AA) such that the ZBL potential has negligible contribution to energies or forces among the initial training configurations where extremely close inter-atomic distances were not sampled. More details about ZBL settings used in this work can be found in Supplementary Information. 

\subsection*{Training Data Generation}

Figure \ref{fig:crystal}(a) shows the hexagonal $P6/mmm$ unit cell of $\alpha$-\ce{Li3N}, where Li2 sites form \ce{Li2N} layers with N sites in the $ab$ plane and Li1 sites connect N sites in neighboring \ce{Li2N} layers along the $c$ axis. To sample a diverse set of configurations, the initial training set includes two major components: 
\begin{enumerate}
    \item Starting from the relaxed $\alpha$-\ce{Li3N} unit cell, we first generated two series of unit cells with lattice distortions. One series samples different lattice constants $a$ and $c$, and the other samples unit cells with different levels of strains (-1\% to 1\% at 0.2\% intervals) applied in six different modes as described in \citet{Jong2015} 
    \item Snapshots were extracted from \textit{ab initio} molecular dynamic (AIMD) simulations at temperatures from 400 K to 1200 K at 200 K intervals under an NVT ensemble. Starting from a $3\times3\times3$ supercell with equilibrium volume, for each temperature, 200 snapshots were taken from a 40 ps AIMD simulation. 
\end{enumerate}

To ensure accurate energies and forces, static DFT calculations were performed on all configurations (including snapshots from AIMD).

\subsection*{Model Training and Test}

\begin{table}[htp]
\centering
\caption{Number of configurations ($N_{conf}$), number of atoms ($N_{atoms}$), and sample weights on energy ($w_E$) and forces ($w_F$) for initial training data used in eSNAP model training. }
\label{tab:fit_weight}
\begin{tabular}{ccccc}
\hline
Type & $N_{conf}$ & $N_{atoms}$ & $w_{E}$ & $w_{F}$ \\
\hline
Distorted unit cells & 109 & 4 & $10^3$ & 0 \\
AIMD snapshots & 1000 & 108 & 1 & $10^{-3}$ \\
\hline
\end{tabular}
\end{table}

Table \ref{tab:fit_weight} shows the weights applied on the different sets of training configurations during model training. As the initial training dataset contains many more configurations from AIMD snapshots with larger number of atoms, a much larger weight was applied on the energies of the distorted unit cells relative to those from the AIMD snapshots. A zero weight was applied on the negligibly small forces for the distorted unit cells. 

\begin{figure}[htp]
\centering
\includegraphics[width=0.5\textwidth]{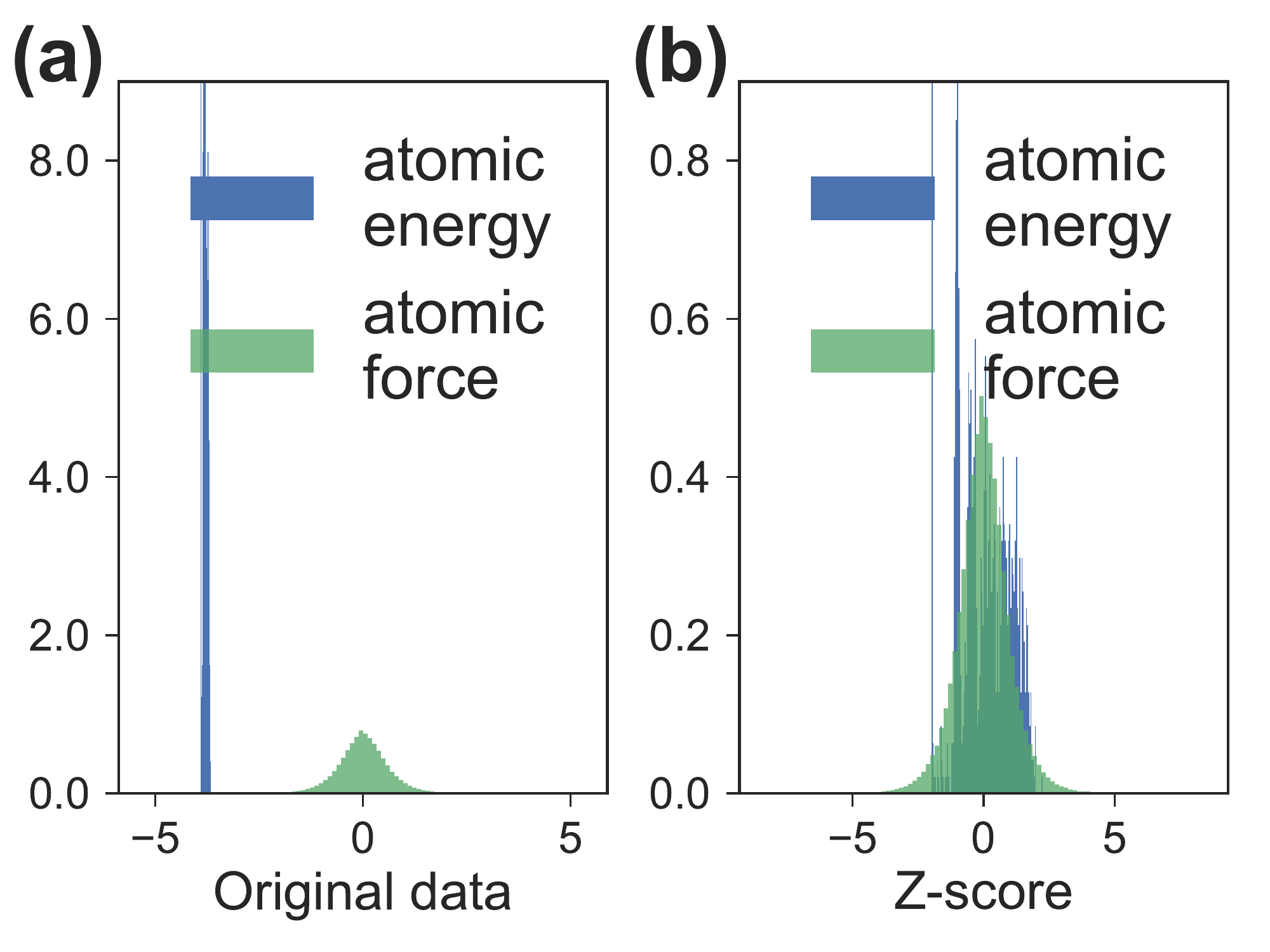}
\caption{Distribution of (a) original atomic energies and forces and (b) normalized z-score of atomic energies and forces.}
\label{fig:z-score}
\end{figure}

As shown in Figure \ref{fig:z-score}a, the energies and forces differ greatly in magnitude and distribution due to differences in the scales and units. In the original SNAP training approach, the effect of this difference in magnitude and distribution is partially accounted for by treating the data weights as hyperparameters to be optimized.\cite{Chen2017b, Li2018c} In this work, we use the standardized z-scores of energies and forces (plotted in Figure \ref{fig:z-score}b) as the targets in model training to avoid incorporating the effect of the distribution in the data weights, which are therefore fixed at the values in Table \ref{tab:fit_weight}. The ``standardized" eSNAP model in the fitting process is then given by the following:
\begin{align}
\begin{bmatrix}
\frac{e - \bar{e}}{\sigma_e} \\ \vdots 
\end{bmatrix} &= 
\frac{1}{N\sigma_e}
\begin{bmatrix}
E_{el} & N_\alpha & \sum^{N_\alpha}_{i=1}B_{1,i} & \dots & \sum^{N_\alpha}_{i=1}B_{k,i} & \dots \\
\vdots & \vdots & \vdots & \dots & \vdots & \dots 
\end{bmatrix}\boldsymbol{\beta}^{\rm T}, \\
\begin{bmatrix}
\frac{\mathbf{F}_j}{\sigma_F} \\ \vdots 
\end{bmatrix} &=
\frac{1}{\sigma_F}
\begin{bmatrix}
-\frac{\partial E_{el}}{\partial \mathbf{r}_j} & 0 & -\sum^{N_\alpha}_{i=1}\frac{\partial B_{1,i}}{\partial \mathbf{r}_j} & \dots & -\sum^{N_\alpha}_{i=1}\frac{\partial B_{k,i}}{\partial \mathbf{r}_j} & \dots \\
\vdots & \vdots & \vdots & \dots & \vdots & \dots 
\end{bmatrix}\boldsymbol{\beta}^{\rm T},
\end{align}
where $e$ is the energy per atom, $\bar{e}$ is the mean of $e$, and $\sigma_e$ and $\sigma_F$ are the standard deviations of $e$ and $F$, respectively. The mean of forces is omitted since it is close to zero. The coefficient vector $\boldsymbol{\beta}^{\rm T}$ to be solved can be written as:
\begin{equation}
\boldsymbol{\beta}^{\rm T} = 
\begin{bmatrix}
\gamma & \beta_{\alpha,0} - \bar{e} & \beta_{\alpha,1} & \dots & \beta_{\alpha,k} & \dots
\end{bmatrix}^{\rm T}.
\end{equation}

For bispectrum coefficient calculations, we used the implementation available in LAMMPS.\cite{Thompson2015a} The two hyperparameters (cutoff distance $R_\alpha$ and atomic weight $w_\alpha$) for each element (Li and N in the case of \ce{Li3N}) were determined using a two-step grid search scheme for the atomic weights and then followed by the cutoff distances. The mean absolute error (MAE) of forces from a linear model trained on the initial training set was chosen as the metric. For the atomic weights, it should be noted that the atomic density in ionic systems is generally higher than that in metallic systems; hence the search of atomic weights was performed in the range where $|w_\alpha| < 1$. Similarly, the search space for cutoff radius was limited to the range where $R_\alpha < 4$ \AA. The results from grid search (Figure \ref{fig:grid}) and the final hyperparameters (Table \ref{tab:param_coeff}) are available in Supplementary Information. 

\begin{figure}[htp]
\centering
\includegraphics[width=0.7\textwidth]{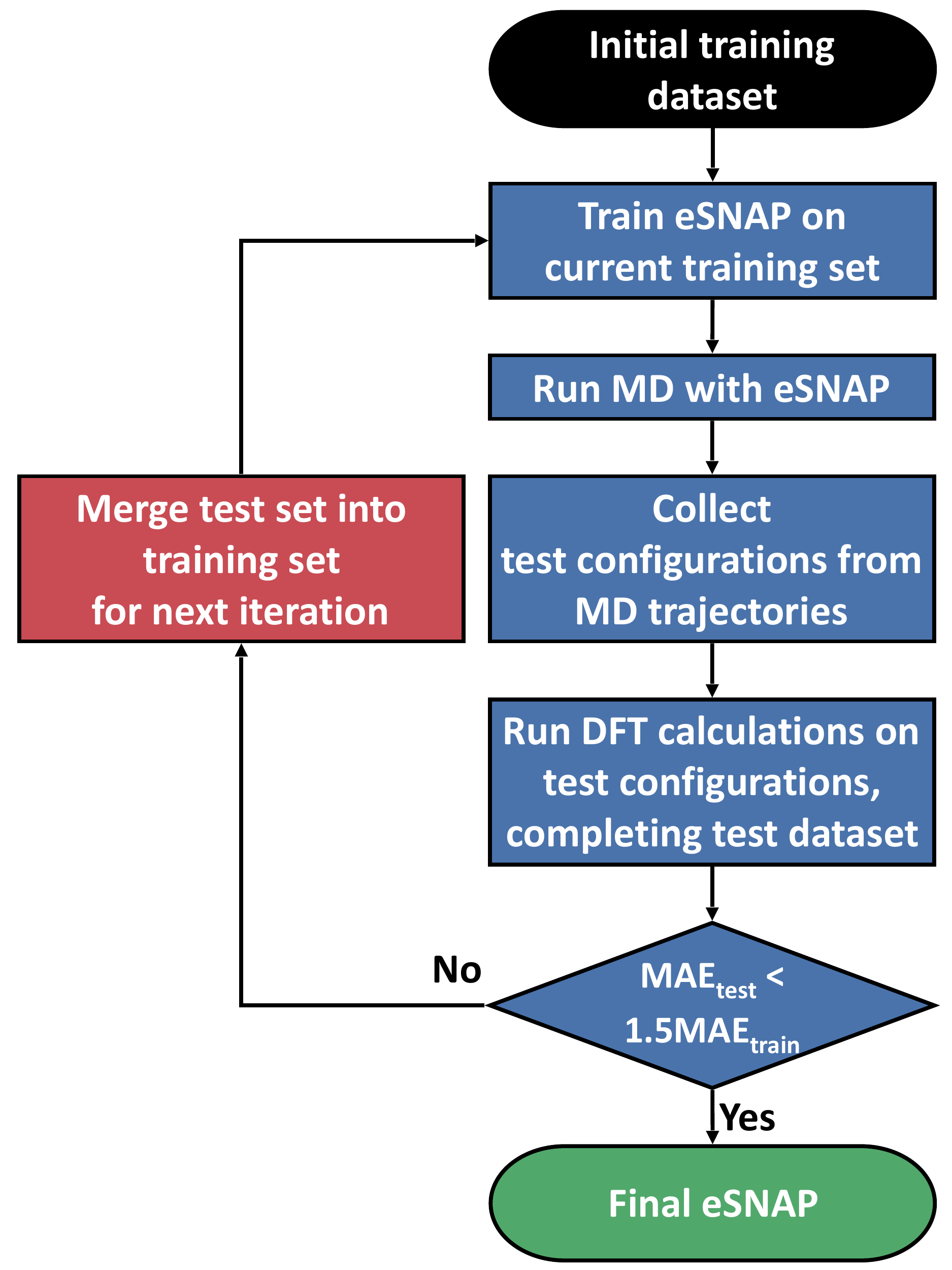}
\caption{\label{fig:flow}Flowchart of iterative procedure for eSNAP model training and test.}
\end{figure}

Figure \ref{fig:flow} show the flow chart of the iterative procedure used for training the eSNAP model in this work. A preliminary eSNAP model was first trained using the initial training set. Using this fitted eSNAP model, MD simulations were then carried out using a $3\times3\times3$ supercell in equilibrium volume at temperatures ranging from 300 K to 1200 K at 100 K intervals under an NVT ensemble for 40 ps. Ten snapshots were sampled from each MD simulation to form a new set of test configurations. Static DFT calculations were performed on these test configurations. If the test MAEs for both energies and forces are significantly larger than the corresponding training MAEs, the test set was then merged into the training set to form a new extended training set. The entire eSNAP fitting, simulation and testing procedure was repeated until there is no significant over-fitting. In this work, we use 150\% of training MAE as the threshold to achieve a balance between the benefit gained by adding more training instances and the associated costs of performing more DFT calculations. It should be noted that this strategy is designed to bias the eSNAP model to improve the predictions on energy and force of MD snapshots, which is the target application of interest in this work.

\subsection*{DFT calculations}

All DFT calculations were performed using the Vienna \textit{Ab initio} Simulation Package (VASP)\cite{Kresse1996} within the projector augmented wave approach.\cite{Blochl1994} The Perdew-Burke-Ernzerhof (PBE) generalized gradient approximation was adopted as the exchange-correlation functional.\cite{Perdew1996a}. To ensure the convergence of energy and atomic force, a plane-wave energy cutoff of 520 eV and $\Gamma$-centered $k$-point meshes with a density of at least 30 \AA\ were employed for all static DFT calculations. For AIMD simulations, a single $\Gamma$ $k$-point and a much lower energy cutoff of 300 eV were used for rapid propagation of trajectories. 

\subsection*{Diffusivity calculations}

The tracer diffusivity of Li $D^*$ is calculated from the mean square displacement (MSD) of all diffusing Li ions as described by the Einstein relation:
\begin{equation}
D^* = \frac{1}{2dt}\frac{1}{N}\sum_{i=1}^N\langle[\Delta\mathbf{r}_i(t)]^2\rangle,
\end{equation}
where $d$ is the number of dimensions in which diffusion occurs, $N$ is the total number of diffusing Li ions, $\Delta\mathbf{r}_i(t)$ is the displacement of the $i$th Li ion at time $t$. 

The charge diffusivity of Li $D_\sigma$ is calculated from the square net displacement of all diffusing Li ions, as described below:
\begin{equation}
D_\sigma = \frac{1}{2dt}\frac{1}{N}\left\langle\left[\sum_{i=1}^N\Delta\mathbf{r}_i(t)\right]^2\right\rangle
\end{equation}

The Li conductivity at temperature $T$ (unit: K) can be calculated from the charge diffusivity $D_\sigma$ using the Nernst-Einstein equation:
\begin{equation}
\sigma = \frac{\rho z^2F^2}{RT}D_\sigma,
\end{equation}
where $\rho$ is the molar density of Li, $z$ is the charge of Li (+1), $F$ is the Faraday constant, and $R$ is the gas constant. 

In addition, the ratio between the tracer and charge diffusivities is referred to in the literature as the Haven
ratio $H_R = D^* / D_\sigma$.

All the simulations with the eSNAP were performed using LAMMPS.\cite{Plimpton1995} All the structure manipulations and interfacing with VASP and LAMMPS were handled by the Python Materials Genomics (pymatgen) library.\cite{Ong2013a}

\section*{Data Availability}

The training configurations and their DFT computed total energy and atomic forces are available in the SNAP development repo on Github (https://github.com/materialsvirtuallab/snap).

\section*{Acknowledgement}

This work was supported by the Office of Naval Research (ONR) Young Investigator Program (YIP) under Award No. N00014-16-1-2621. The authors acknowledge computational resources provided by Triton Shared Computing Cluster (TSCC) at the University of California, San Diego, the National Energy Research Scientific Computing Center (NERSC), and the Extreme Science and Engineering Discovery Environment (XSEDE) supported by the National Science Foundation under Grant No. ACI-1053575. The authors also thank Dr. Anton Van der Ven for the discussion on diffusivity calculations.

\section*{Author Contributions}

Z.D. performed potential model training, performance evaluation and diffusion studies. C.C. and X.-G. L. helped with the design of training procedure and analyses in diffusion studies. S.P.O. is the primary investigator and supervised the entire project. All authors contributed to the writing and editing of the manuscript.

\section*{Competing Interests Statement}

The authors declare no competing interests.

\end{document}


\section*{Nuclear repulsion}
The Ziegler-Biersack-Littmark (ZBL) screened nuclear repulsion between a pair of atoms $i$ and $j$ at a distance $r_{ij}$ is given by:
\begin{align*}
E^{ZBL}_{ij} &= \frac{1}{4\pi\epsilon_0}\frac{Z_iZ_je^2}{r_{ij}}\phi(r_{ij}/a) + S(r_{ij}) \\
a &= \frac{0.46850}{Z^{0.23}_i + Z^{0.23}_j} \\
\phi(x) &= 0.18175e^{-3.19980x} + 0.50986e^{-0.94229x} + 0.28022e^{-0.40290x} + 0.02817e^{-0.20162x}
\end{align*}
where $e$ is the electron charge, $\epsilon_0$ is the electrical permittivity of vacuum, and $Z_i$ and $Z_j$ are the nuclear charges of the two atoms. $S(r)$ is a switching function that ramps the energy, force, and curvature smoothly to zero between inner ($R_i$) and outer ($R_o$) cutoff.

\clearpage

\section*{Figures}

\begin{figure}[h]
\subfloat[]{
\includegraphics[width=0.6\textwidth]{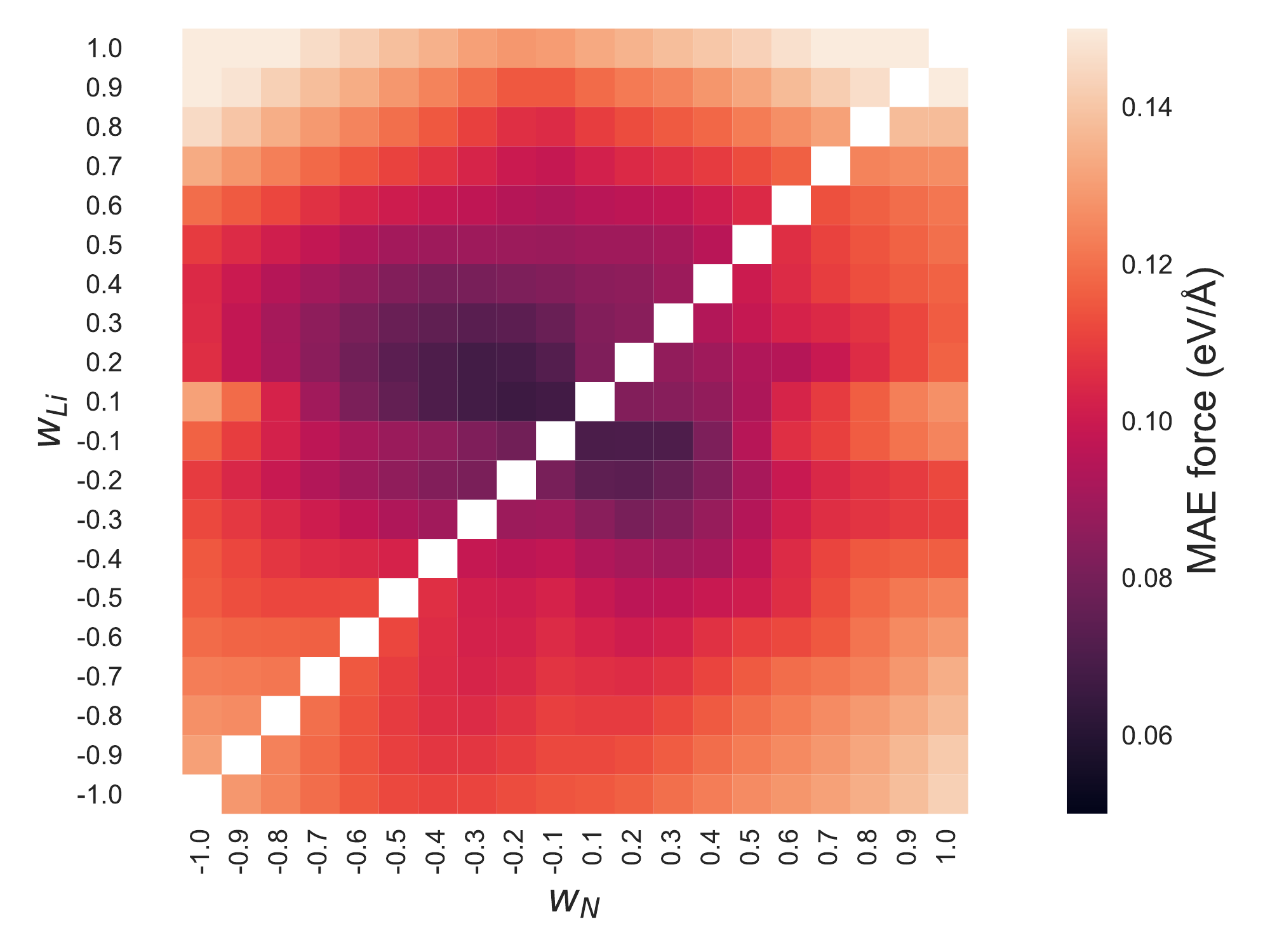}
}
\quad
\subfloat[]{
\includegraphics[width=0.6\textwidth]{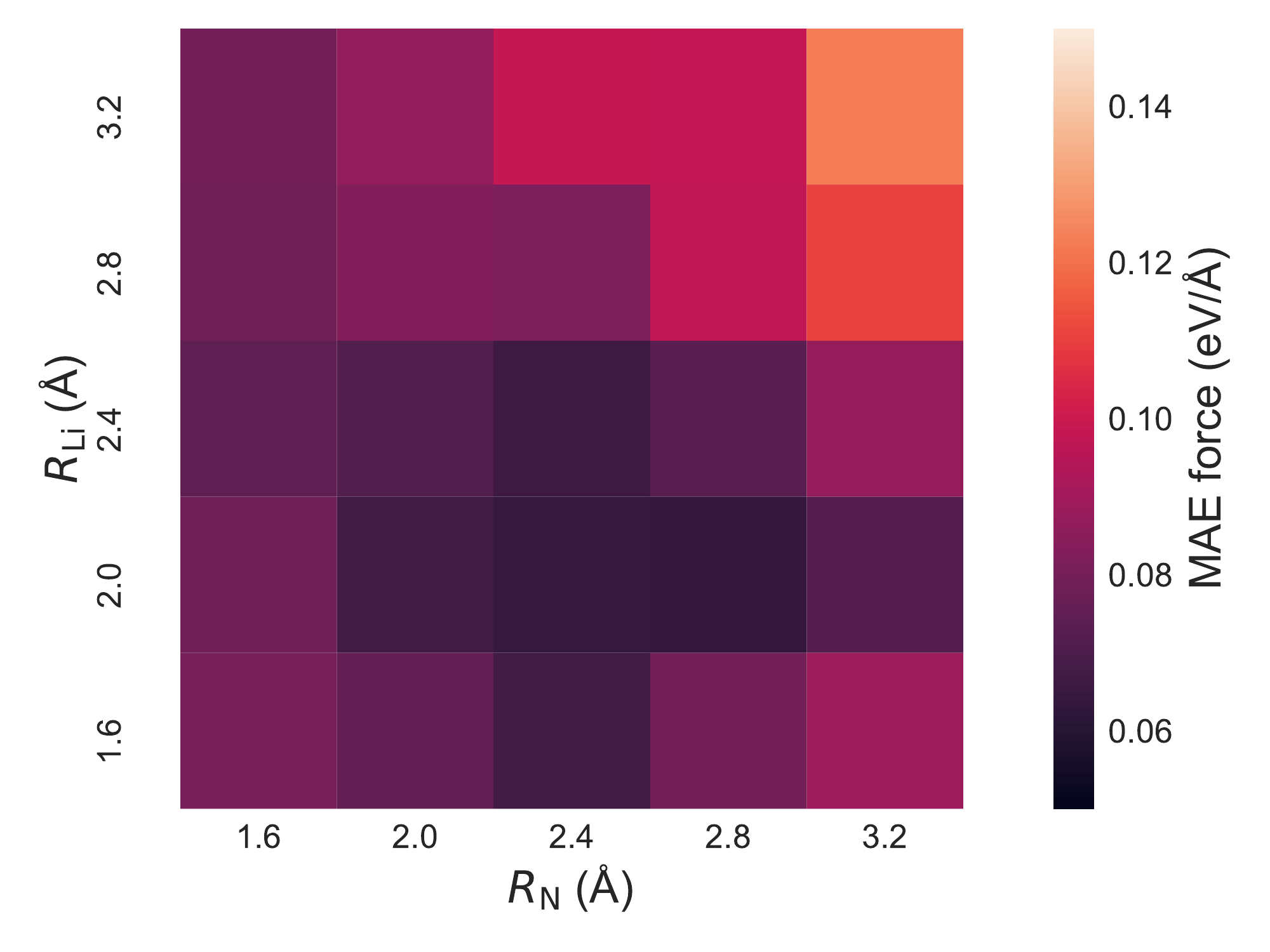}
}
\caption{Grid search of (a) atomic weights $w_\alpha$ and (b) cutoff distances $R_\alpha$. Mean absolute error on forces is used as the metric.}
\label{fig:grid}
\end{figure}

\begin{figure}[h]
\includegraphics[width=0.6\textwidth]{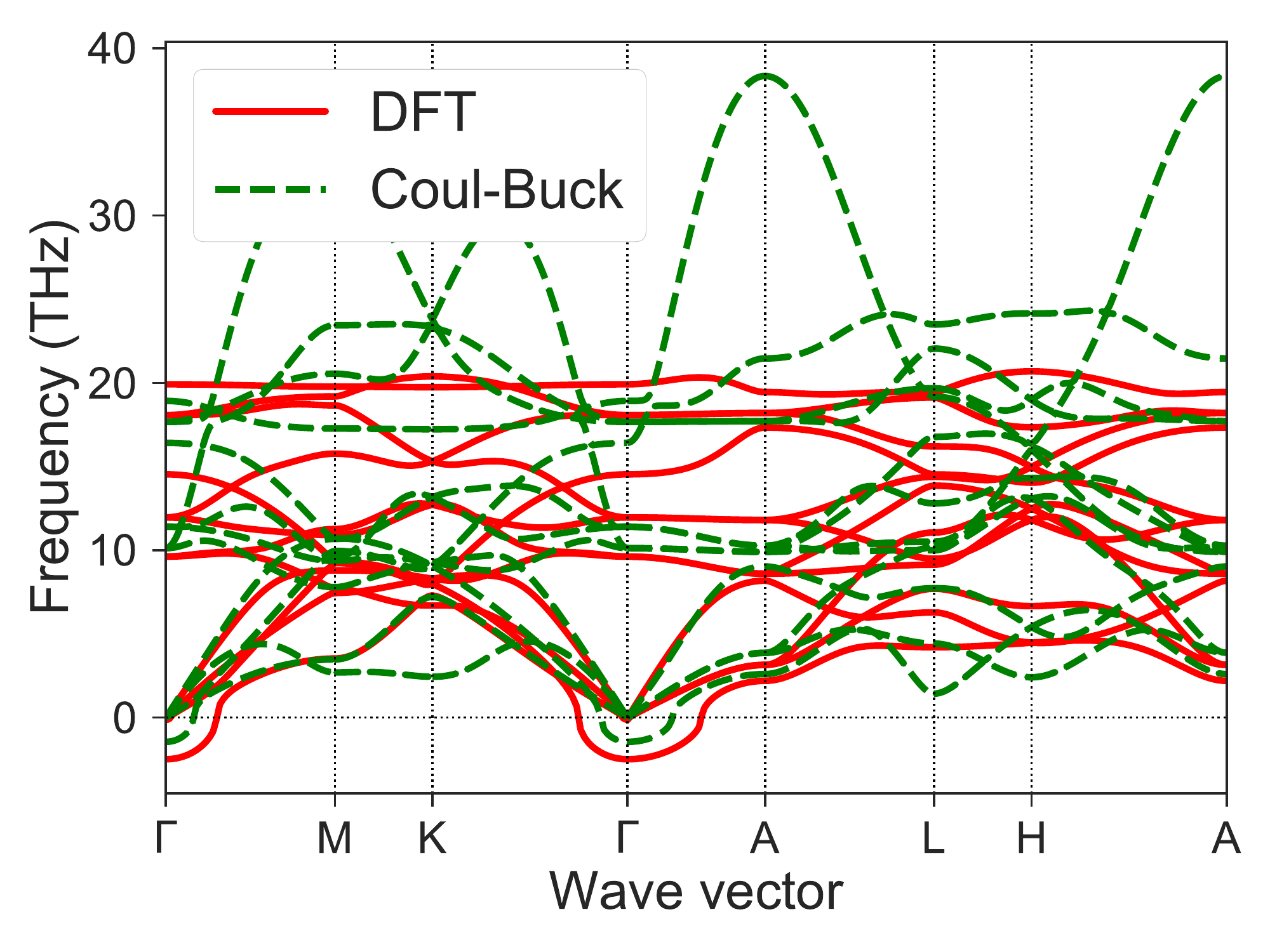}
\caption{Phonon dispersion curves of $\alpha$-\ce{Li3N} calculated from Coulomb-Buckingham potential in comparison with DFT.}
\label{fig:phonon_buck}
\end{figure}

\begin{figure}[h]
\includegraphics[width=0.6\textwidth]{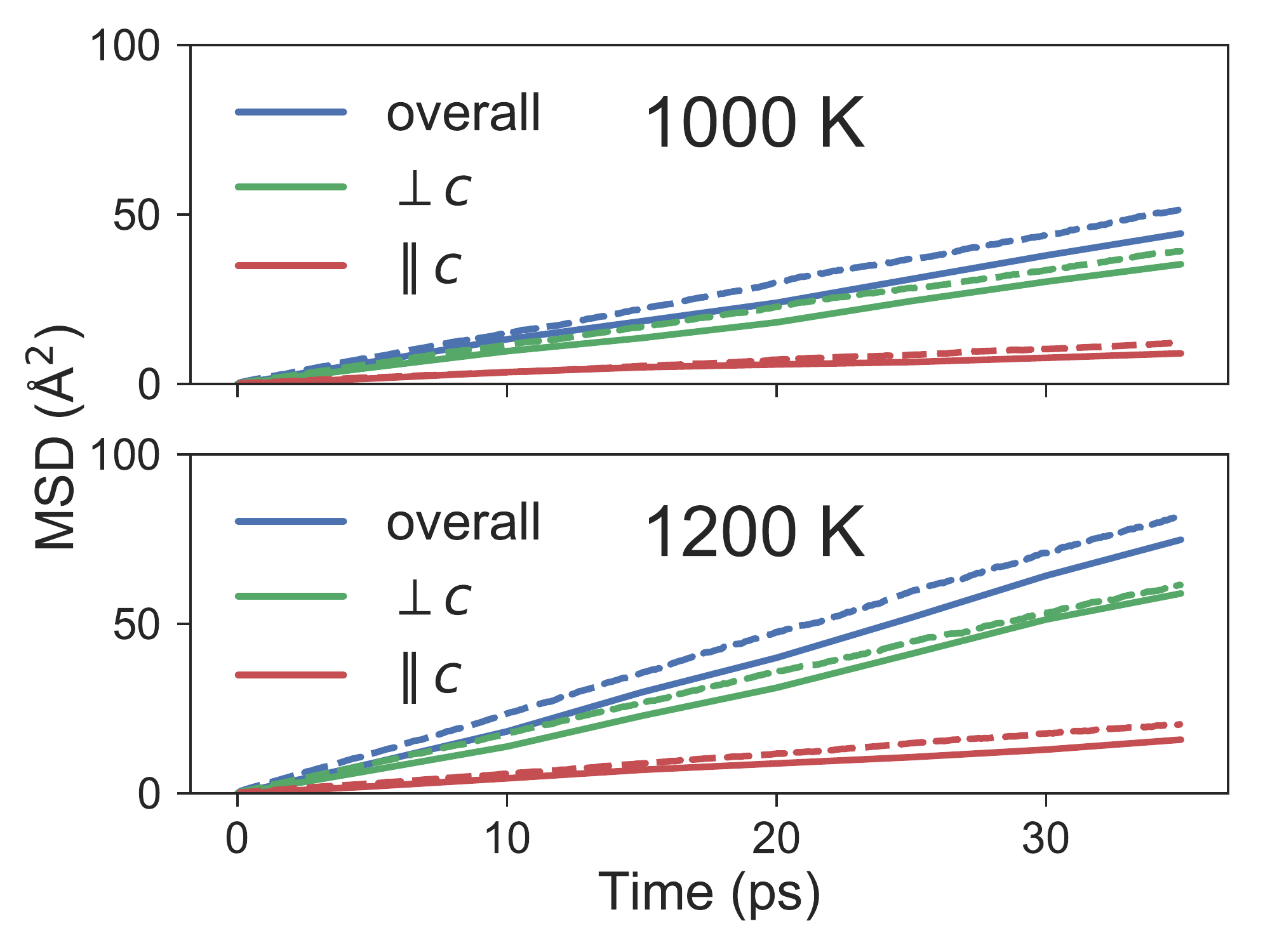}
\caption{MSD vs. time plots for simulations running with AIMD (solid lines, smoothed) and eSNAP (dashed lines) at 1000 (upper) and 1200 (lower) K.}
\label{fig:msd_bulk}
\end{figure}

\begin{figure}[h]
\includegraphics[width=0.6\textwidth]{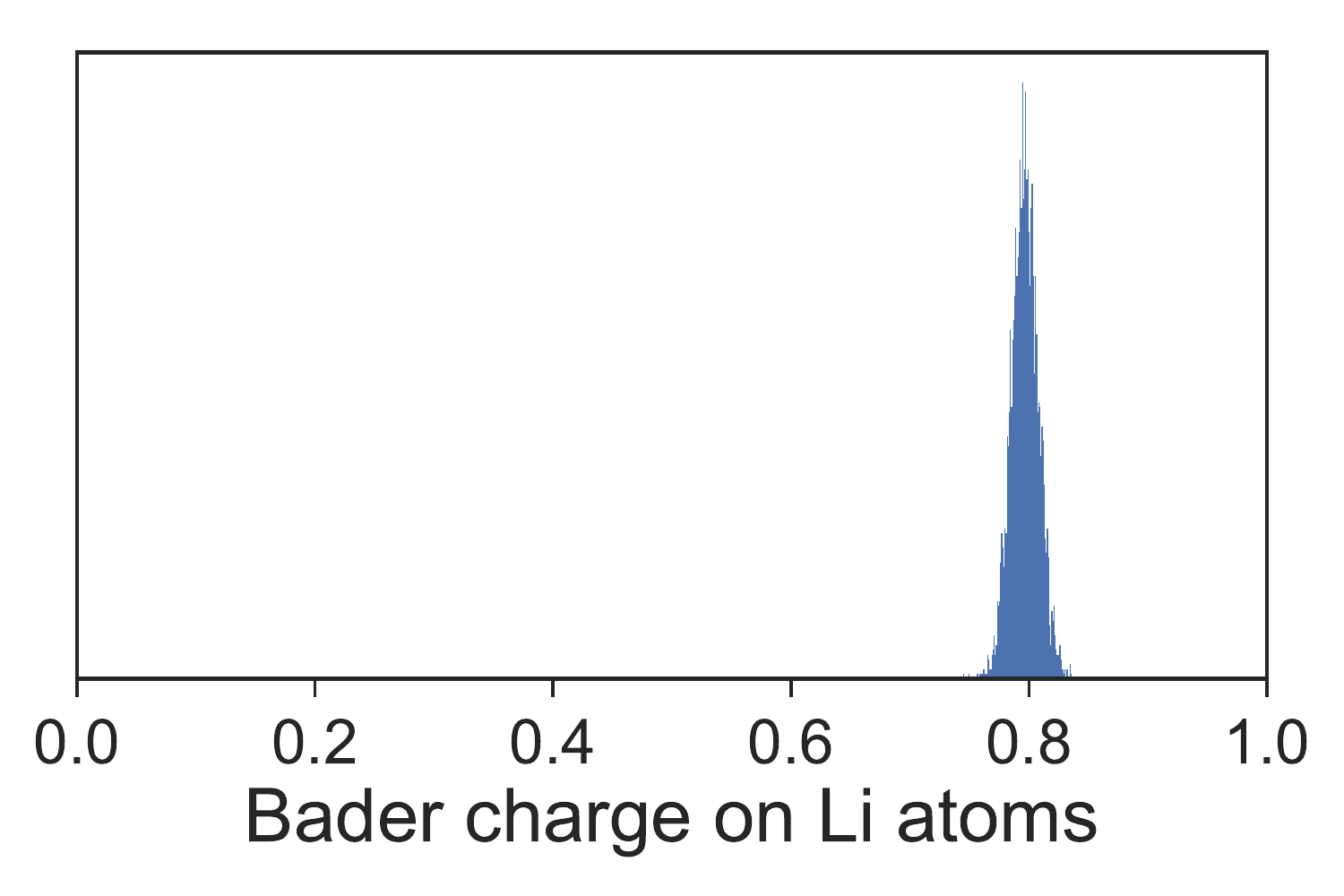}
\caption{Bader charge distribution for Li atoms in the initial training data structures.}
\label{fig:bader}
\end{figure}

\begin{figure}[h]
\includegraphics[width=0.6\textwidth]{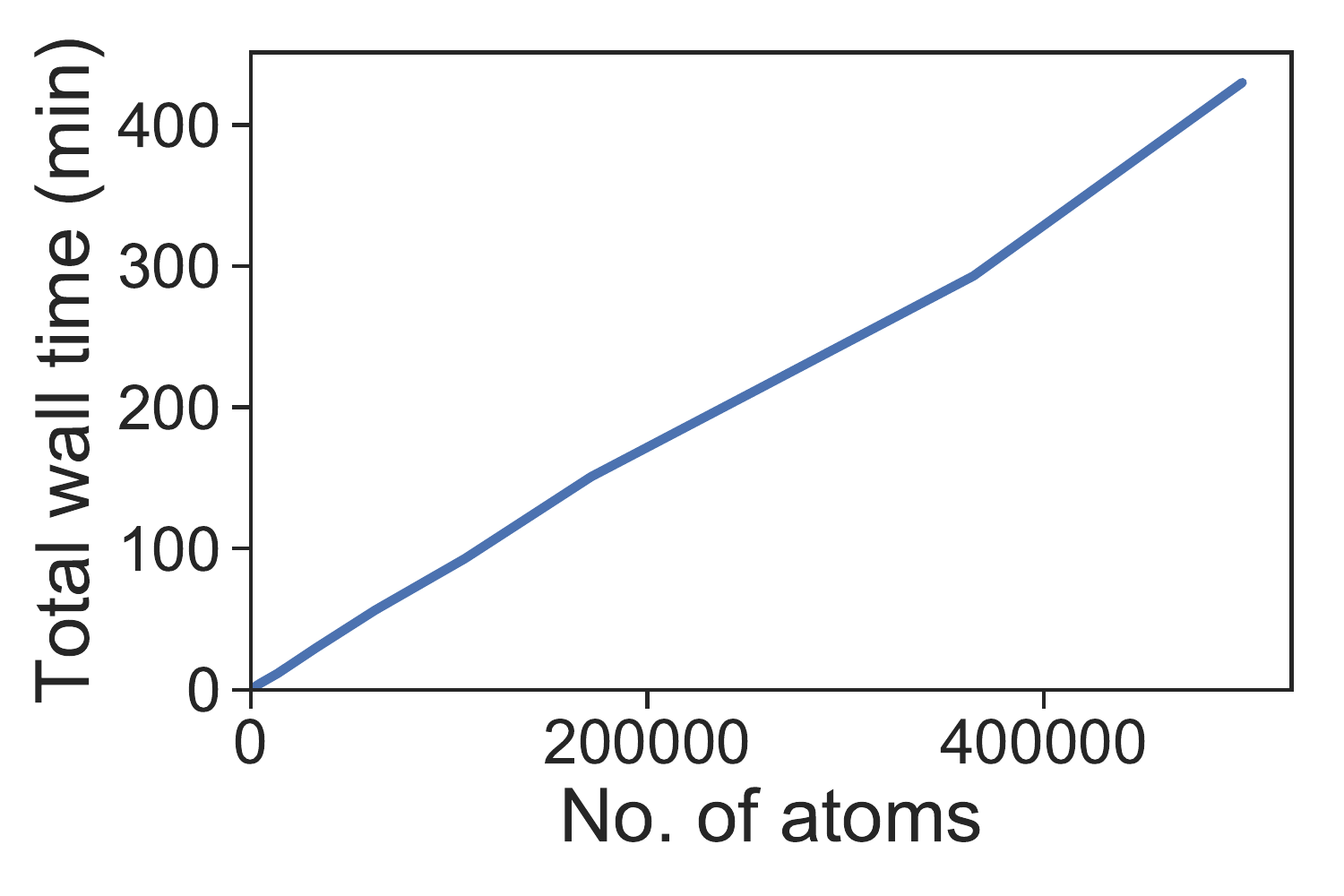}
\caption{Scaling performance of eSNAP tested on 12 CPUs of 1 Intel Haswell Compute node.}
\label{fig:scale}
\end{figure}